\documentclass[twocolumn,aps,prl,superscriptaddress]{revtex4-2}
\usepackage{amsmath,empheq}
\usepackage{amssymb}
\usepackage{mathrsfs}  
\usepackage{amsfonts}
\usepackage{booktabs}
\usepackage{physics}
\usepackage{graphicx} 
\usepackage{subfigure} 
\usepackage{epsfig}
\usepackage{color}
\usepackage{epstopdf}
\usepackage{xspace}
 \usepackage{rotating}
 \usepackage{appendix}
 \usepackage{chemformula}
 \usepackage{bm}
 \usepackage[colorlinks=true,linkcolor=blue,citecolor=blue,filecolor=blue,urlcolor=blue]{hyperref}

 \definecolor{bondiblue}{rgb}{0.0, 0.58, 0.71}

\begin{document}

\title{Emergence of Spinon Fermi Arcs in the Weyl-Mott Metal-Insulator Transition}
\author{Manuel Fern\'andez L\'opez}
\affiliation{Departamento de F\'isica Te\'orica de la Materia Condensada, Condensed Matter Physics Center (IFIMAC) and
Instituto Nicol\'as Cabrera, Universidad Aut\'onoma de Madrid, Madrid 28049, Spain}
\author{Iñaki García-Elcano}
\affiliation{Departamento de F\'isica Te\'orica de la Materia Condensada, Condensed Matter Physics Center (IFIMAC) and
Instituto Nicol\'as Cabrera, Universidad Aut\'onoma de Madrid, Madrid 28049, Spain}
\author{Jorge Bravo-Abad}
\affiliation{Departamento de F\'isica Te\'orica de la Materia Condensada, Condensed Matter Physics Center (IFIMAC) and
Instituto Nicol\'as Cabrera, Universidad Aut\'onoma de Madrid, Madrid 28049, Spain}
\author{Jaime Merino}
\affiliation{Departamento de F\'isica Te\'orica de la Materia Condensada, Condensed Matter Physics Center (IFIMAC) and
Instituto Nicol\'as Cabrera, Universidad Aut\'onoma de Madrid, Madrid 28049, Spain}

\begin{abstract}
The Weyl-Mott insulator (WMI) has been postulated as a novel type of correlated insulator with non-trivial topological properties. We introduce a minimal microscopic model that captures generic features of the WMI transition in Weyl semimetals. The model hosts a bulk Mott insulator with spinon Fermi arcs on its surfaces which we identify as a WMI. At finite temperatures, we find an intermediate Weyl semimetallic phase with no quasiparticles which is consistent with the bad semimetallic behavior observed in pyrochlore iridates, A$_2$Ir$_2$O$_7$, close to the Mott transition. Spinon Fermi arcs lead to a suppression of the bulk Mott gap at the surface of the WMI, in contrast to the gap enhancement found in conventional Mott insulators, which can be detected through angular resolved photoemission spectroscopy (ARPES). 
\end{abstract}
 \date{\today}
 \maketitle
Weyl fermions are under intense research in condensed matter since their theoretical prediction \cite{Savrasov2011,Burkov2011} and subsequent observation of topological Fermi arcs in TaAs
materials \cite{Hassan2015,Lv2015}. In contrast, strongly correlated phenomena such as the Mott insulator transition in Weyl semimetals (WSM) have been much less explored \cite{Yang2011,Hosur2012,Rosenstein2013,Balents2014,Savary2014}. There are open fundamental 
questions regarding the WSM to Mott insulator transition:
is the Mott insulator arising from the WSM conventional or does it retain topological properties of the Weyl fermions and, which are the key electronic properties of a WSM around the Mott transition? 
These questions are relevant to strongly interacting WSM with spin-orbit coupling (SOC) such as A$_2$Ir$_2$O$_7$, to artificial systems~\cite{Wang2021,Lu2020,Goikoetxea2020} in which interactions can be tuned in a controllable way, and to photonic environments doped with atom-like emitters~\cite{Noh2017,Innaki2020,Innaki2021,Innaki2023}.
 
Strikingly, the Weyl-Mott insulator (WMI) has been predicted to be topological, hosting {\it surface Fermi arcs concomitantly with a Mott bulk gap} \cite{Nagaosa2016}. However, the theoretical prediction of the WMI is based in a rather unrealistic model which considers a local Coulomb interaction in momentum space instead of the on-site Hubbard $U$ relevant to actual materials.  Conversely, dynamical mean-field theory \cite{Hofstetter2021} (DMFT) on a Weyl-Hubbard model of cold atoms in optical lattices \cite{Buljan2015} finds a transition from a WSM to a topologically trivial Mott insulator in contrast to the WMI. Since DMFT only includes local electron correlation effects associated with $U$ exactly, 
an important question is whether non-local correlations can lead to topological surface bands traversing the Mott gap. This is one key issue addressed here. 

\begin{figure}[t!]
    \centering
    \includegraphics[width=7.5cm,clip]{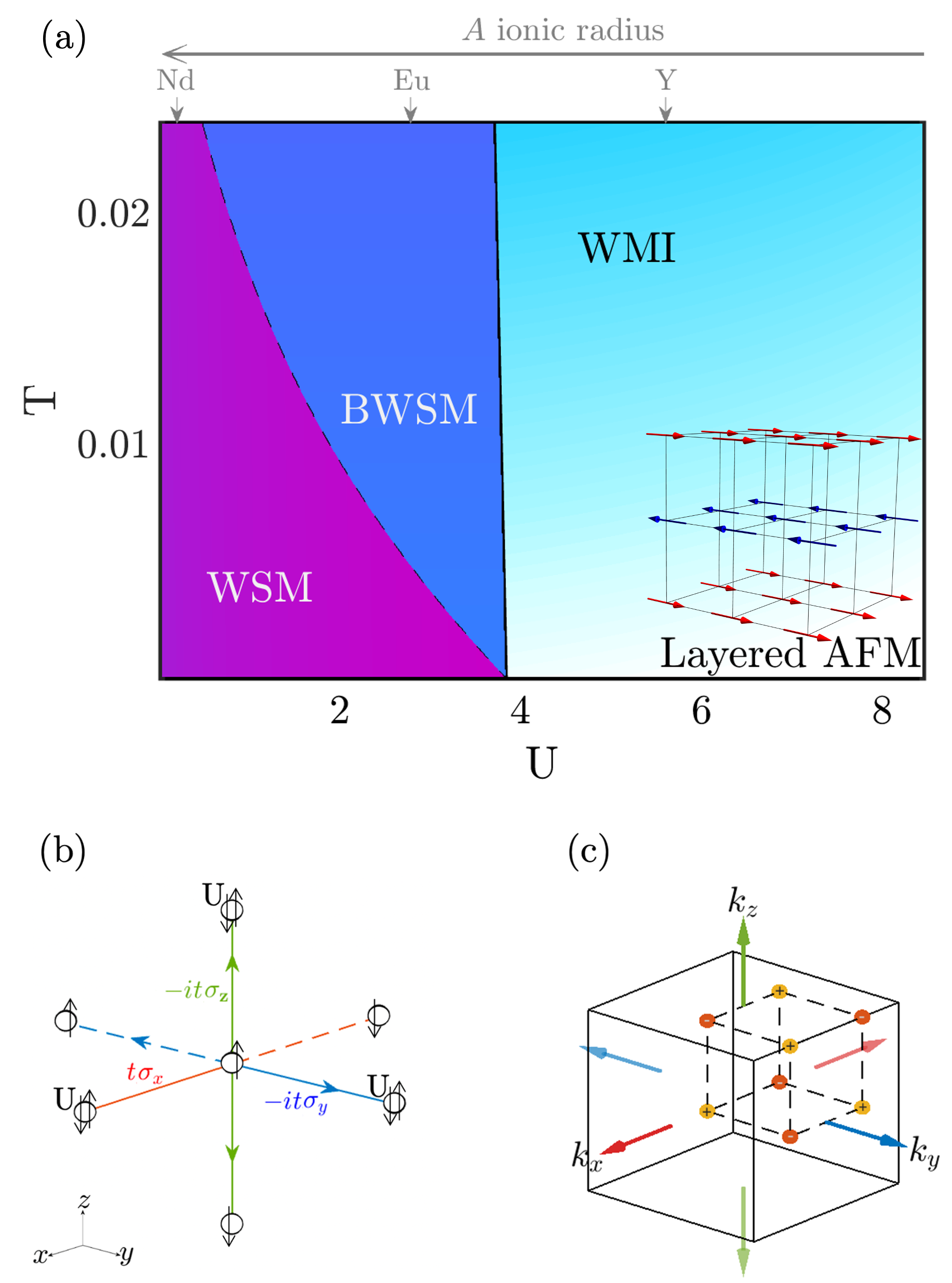} 
    \caption{(a) Temperature vs. $U$ phase diagram of the Weyl-Mott metal-insulator transition. The WMI characterized by a finite charge gap and spinon Fermi arcs on the surface occurs at large $U$. 
    Layered AFM order arises at sufficiently low $T$ in the WMI. Increasing $U$ corresponds to decreasing the radius of the $A$-ion in the pyrochlore iridates, A$_2$Ir$_2$O$_7$. (b) Real space illustration of our Weyl-Hubbard model (\ref{eq:Hubbard}) describing the Weyl-Mott metal-insulator transition. (c) Location of the spinon Weyl points in the 1st Brillouin zone of the cubic lattice. Positive (negative) topological charges are indicated as yellow (orange) spheres.  
    We have fixed $t \equiv 1$.}
\label{fig:phased}
\end{figure}

The characterization of the Weyl-Mott insulator transition is relevant to the pyrochlore iridates, A$_2$Ir$_2$O$_7$. As the volume of the $A$ ion increases a metal-to-magnetic insulator occurs \cite{Balents2014}. While a mean-field approach predicts an intermediate AF ordered WSM phase \cite{Savrasov2011,Balents2014}, which has not yet been observed, a DMFT calculation \cite{Werner2015} predicts a direct metal to AF ordered Mott insulator transition. AF fluctuations around a novel quantum critical point can lead to anomalous exponents \cite{Savary2014}. At finite $T$, semiconducting ($A=Lu,Yb,Ho,Y, Dy$), 
semimetallic
($A=Gd,Eu, Sm$), and metallic ($A=Nd,Pr$) phases are observed which remain poorly understood \cite{Li2021}.

In the present Letter, we show how a WMI having topological spinon surface states traversing the Mott-Hubbard gap emerges in a model of a strongly interacting WSM breaking time-reversal symmetry. Spinon Fermi arcs lead to the suppression of the Mott gap at the surface of the WMI which can be detected in ARPES. Such behavior is in contrast with the gap enhancement encountered at the surface of a conventional Mott insulator not hosting spinon Fermi arcs. 
The WSM, 'bad' WSM (BWSM) and WMI arising in our phase diagram (shown in Fig.~\ref{fig:phased}(a)) are in one-to-correspondence with the three phases observed in A$_2$Ir$_2$O$_7$ with decreasing the $A$ radius across the Mott insulator transition.

We introduce a simple microscopic model which captures 
the competition between the kinetic energy and on-site Coulomb repulsion in a 
WSM:
\begin{align}
\mathcal{H}= \mathcal{H}_0 +\mathcal{H}_U=t\sum_{{\langle ij\rangle_a}} e^{i\phi_a}c_{i\alpha}^\dagger \sigma_{\alpha\beta}^a c_{j\beta}+\frac{U}{2}\sum_i (n_i-1)^2,
\label{eq:Hubbard}
\end{align}
where the first term, $\mathcal{H}_0$,
is the non-interacting part with $\sigma^a$ the Pauli matrices, $\phi_a=0,-\pi/2,-\pi/2$ for $a=x,y,z$ and $\alpha,\beta=\uparrow,\downarrow$. The hopping structure encoded in $\mathcal{H}_0$ illustrated in Fig. \ref{fig:phased}(b), describes a WSM on a simple cubic lattice with broken time-reversal symmetry. The corresponding hamiltonian in momentum space is a two-band model~\cite{Yang2011}: 
$\mathcal{H}_0({\bf k})=2t (\cos(k_xa)\sigma_x +\sin(k_ya)\sigma_y+\sin(k_za)\sigma_z)$
containing 8 Weyl nodes shown in Fig.\ref{fig:phased}(c).
Hence we study an ideal WSM described by $\mathcal{H}_0$ which have been artificially generated through cold atoms in optical lattices \cite{Wang2021,Lu2020}. 
\begin{figure}[t!]
    \centering
   \includegraphics[width=8.5cm,clip]{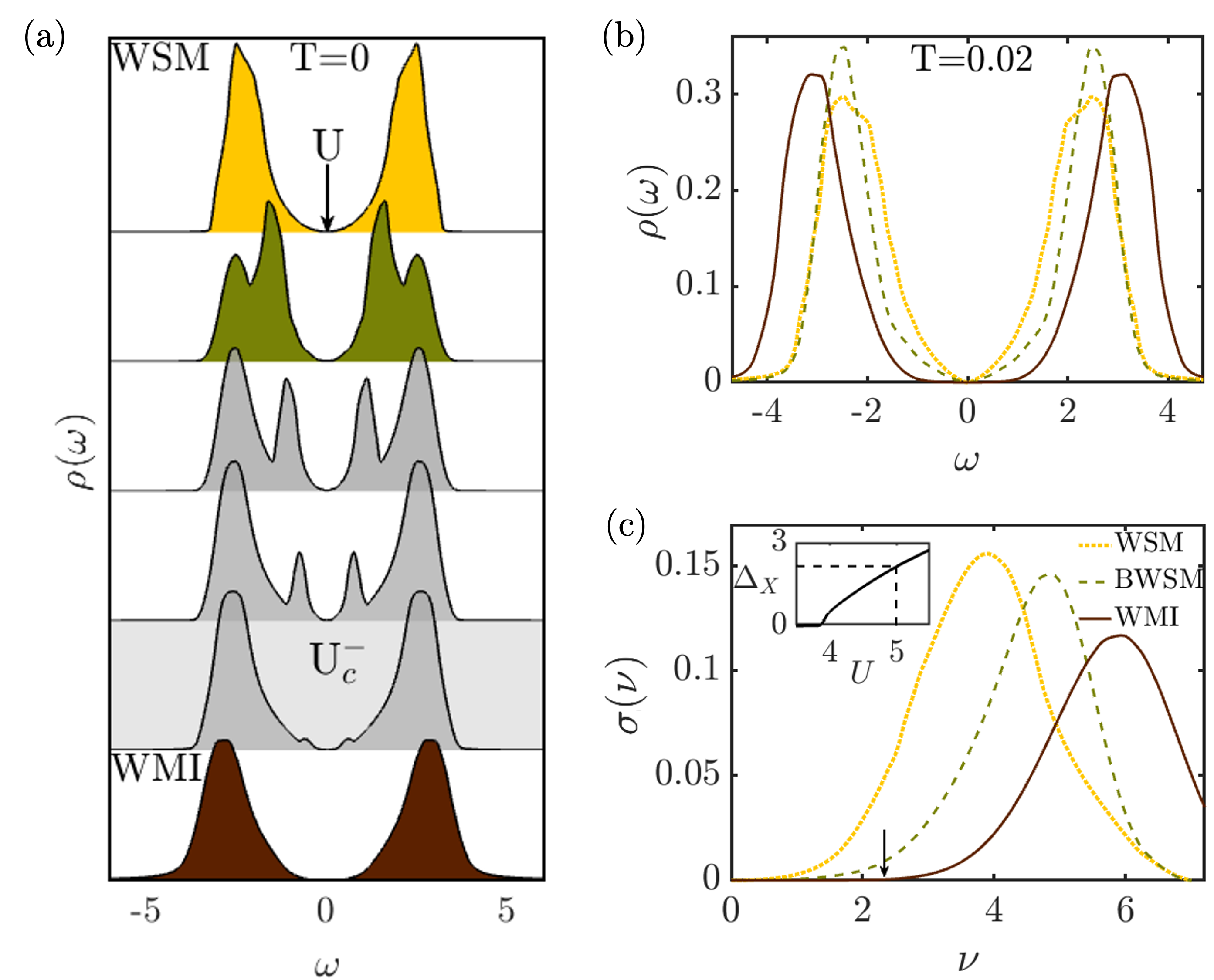} 
    \caption{(a) Dependence of the zero temperature electron DOS across the Weyl-Mott metal-insulator transition for $U=0^+,1,2,3,U_c^-,5$. Approaching the Mott transition $U \rightarrow U_c=3.84$, the quasiparticle weight $Z \rightarrow 0$ and the spinon bandwidth narrows due to the suppression of $Q_f$. Hubbard bands separated by a charge gap $\Delta_X$ determine the DOS of the WMI. 
    (b) Finite temperature DOS across the WMI transition. The DOS of the non-interacting ($U=0$) WSM is compared with that of the BWSM and WMI at a finite $T=0.02$. (c) Optical conductivity for the corresponding cases shown in (b). Inset: Dependence of the charge gap, $\Delta_X$, on $U$. The vertical arrow denotes $\Delta_X$ in the WMI. We have fixed $t \equiv 1$.}
    \label{fig:MIT}
\end{figure}

The Weyl-Hubbard model~(\ref{eq:Hubbard}) is solved using slave-rotor mean-field theory (SRMFT) \cite{Florens2002,Florens2004,Pesin2010,Rachel2010,Jana2019,Du2020,Ko2011,Sorn2018,Jing2011,Young2008,Balents2023}. The slave rotor approach splits the electron into a neutral fermion carrying only spin (spinon) and a boson carrying only charge (rotor): $c_{i\sigma}^\dagger =f_{i\sigma}^\dagger X_i$. Introducing such partitioning in model~(\ref{eq:Hubbard}) and after performing a SRMFT decoupling, the hamiltonian reads:

\begin{align}
\mathcal{H}_X=& -t\sum_{\langle ij\rangle_a}Q_X^a X_i^*X_j
+\sum_i \left( \frac{U}{2} L_i^2 + \lambda  X_i^*X_i\right),\\
\mathcal{H}_f=& t\sum_{{\langle ij\rangle_a}} Q_f^ae^{i\phi_a}\ f_{i\alpha}^\dagger \sigma_{\alpha\beta}^af_{j\beta},
\end{align}
where $Q_X^a=\sum_{\alpha\beta}\langle f_{i\alpha}^\dagger\sigma_{\alpha\beta}^af_{j\beta}\rangle_a$ and $Q_f^a=\langle X_i^*X_j\rangle_a$ are the renormalization factors due to $U$, $\lambda$ is the Lagrange multiplier which imposes the charge conservation and $L_i$ is de charge angular momentum.
Hence, the Weyl-Hubbard model  for the original electrons is converted into a quantum XY model for the rotors renormalized by $Q_X$ coupled to 
spinons moving in the original lattice with kinetic energies renormalized by $Q_f$. 
Another two essential parameters for the Mott transition characterization are: the charge gap $\Delta_X$; and the quasiparticle weight $Z$, {\it i. e.}, the coherent fraction of the electron surviving Coulomb interaction effects \cite{suppl}.

\begin{figure*}[t!]
    \centering
    \includegraphics[width=12.5cm,clip]{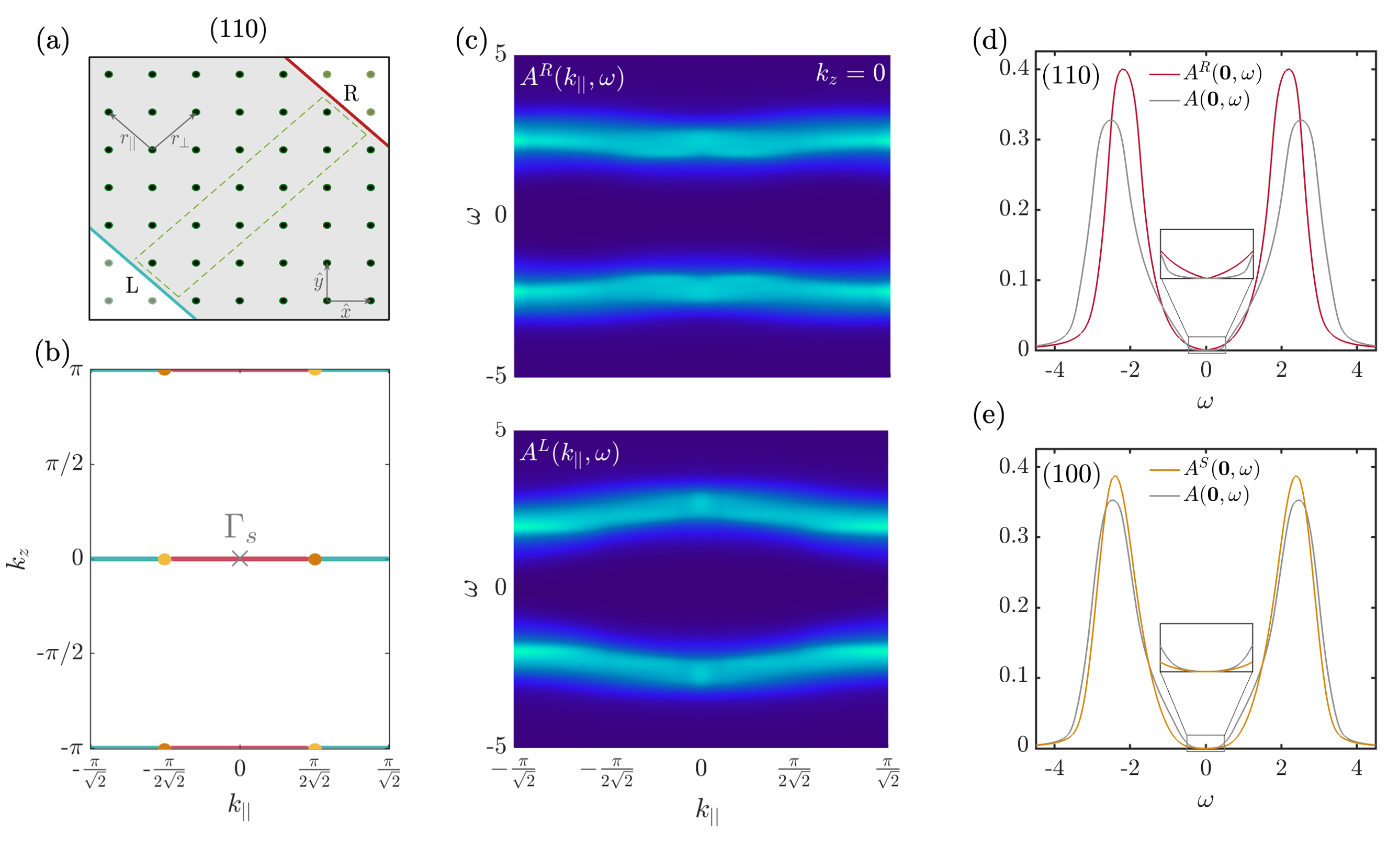}
    \caption{Signatures of spinon Fermi arcs on the spectral functions of a WMI. (a) Projection in $x-y$ plane of a slab cut along the $(110)$ direction of the crystal. (b) Spinon Fermi 'arcs' connecting the projected Weyl points on the $(110)$ surface Brillouin zone on the left (red) and right (cyan) surfaces of the slab.  (c) Spectral function, $A^{R/L}(k_{||}, \omega)$, at the $R/L$ faces of the $(110)$ slab. (d) Suppression of the bulk Mott gap at the $(110)$ surface due to the spinon Fermi arcs. The gapless spectral function at the right surface $A^R({\bf k}_s={\Gamma}_s,\omega)$ is compared with the bulk {\it i. e.} at the middle layer of the slab, $A({\bf k}_s={\bf 0},\omega)$, displaying the Mott gap. (e) On the $(100)$ surface, the Mott gap is enhanced at the surface. Spin-up spectral functions are depicted in panels (c)-(e). We have fixed $U=3.9$ and $T=0.01$.
    }
    \label{fig:Fermiarcs}
\end{figure*}

The Weyl-Mott metal-insulator transition occurs at a critical $U=U_c \approx 3.84$ (see Fig. \ref{fig:phased}). As $U$ is increased, the quasiparticle 
weight $Z$ 
is gradually suppressed leading to a correlated WSM which eventually becomes a Mott insulator, with $Z=0$ for $U\leq U_c$. Concomitantly, the rotors become gapped signalling the opening of the charge gap $\Delta_X$ as expected in a conventional Mott insulator. However, in contrast to the behavior of $Z$ the spinon renormalization factor $Q_f$ remains non-zero even in the Mott insulating phase \cite{Florens2004}, implying the existence of gapless spinon excitations. This is due to the non-local short range magnetic correlations which are missed by DMFT calculations predicting, indeed, a much larger $U_c \approx 15$ \cite{Hofstetter2021}. Hence, instead of the topologically trivial paramagnetic Mott insulator obtained with DMFT~\cite{Hofstetter2021,Werner2015} our Mott insulator is a $U(1)$ spin liquid with Weyl nodes in the spinon sector. Due to the bulk-boundary correspondence, topological edge states of spinons leading to Fermi arcs emerge (as  shown in Fig.~\ref{fig:Fermiarcs}). We identify such Mott insulator with spinon Fermi arcs as the WMI.

At finite temperatures, an unconventional semimetallic phase, the BWSM, 
arises between the WSM and WMI. In contrast to the conventional WSM, the BWSM is gapless, $\Delta_X=0$, {\it but} with no quasiparticles, {\it i. e.}, $Z=0$. The three finite-$T$ phases found
are consistent with the metallic, semimetallic and semiconducting phases observed in the experimental phase diagram \cite{Li2021} of A$_2$Ir$_3$O$_7$ with decreasing A radius. We conclude that while the A=Pr perovskite is a correlated WSM, A=Eu is a BWSM and A=Y a WMI (see Fig. \ref{fig:phased}(a)).  
\begin{figure}[t!]
    \centering
    \includegraphics[width=9cm,clip]{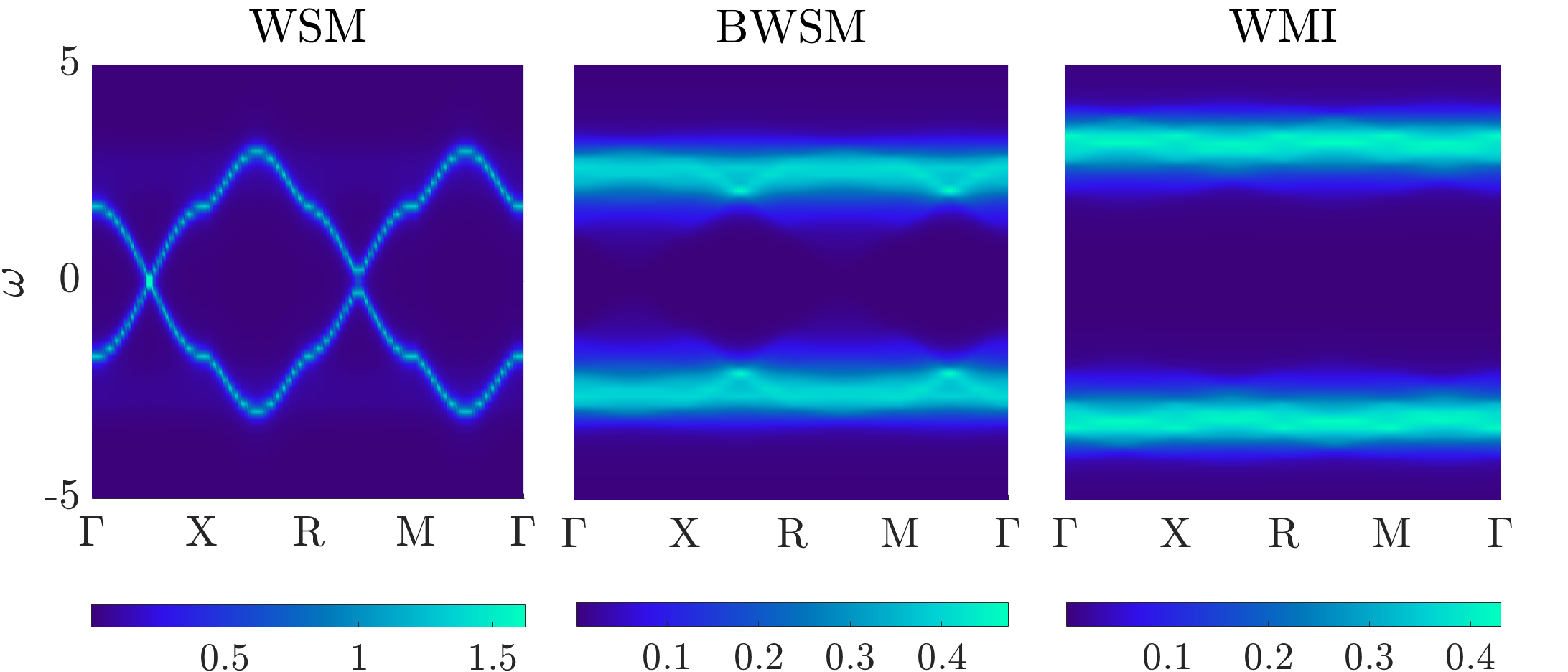}
    \caption{Momentum resolved spin averaged electron spectral function, $A({\bf k},\omega)$, across the WMI transition at a finite temperature. In the WSM (left panel) $A({\bf k},\omega)$ displays coherent quasiparticles which disappear in the BWSM (center panel) although the system remains metallic. The WMI (right panel) 
    displays Hubbard bands centered at $\omega \sim \pm U/2$ with a charge gap $\Delta_X$. 
    The parameters in the WSM, BSWM, and WMI are $U=0.1, 1.7$ and $U=7$ at $T=0.02$.
    }
    \label{fig:spectral}
\end{figure}

Transport in BWSMs is incoherent, meaning that resistivities 
violate the Mott-Ioffe-Regel (MIR) limit \cite{Hussey2004}. In undoped semi-metals, such limit 
implies: $k_T l = 2\pi$, where $k_T= k_B T / \hbar v$ is the wave-vector of thermally generated fermions \cite{Ramirez2020}, $l$ the electron mean-free path and $v$ the electron velocity at the Weyl node. Indeed, the absolute values of the resistivities of A=Eu \cite{Krempa2014,OFarrell2012} and Nd \cite{Ueda2012} materials above $T_N$ display semimetallic behavior with absolute values violating the Mott-Ioffe-Regel limit, $k_T l < 2 \pi$. \cite{suppl} In contrast the resistivities of A=Pr satisfy $k_T l > 2 \pi$ which is consistent with a weakly correlated WSM. In the insulating side of the transition, the $Y$-salt strongly violates the MIR limit \cite{Ramirez2020}, $k_T l \ll 2 \pi$.

The zero temperature Weyl-Mott metal insulator transition can be further characterized by the behavior of the density of states (DOS) $\rho(\omega)$ with $U$~ \cite{suppl}.
The $U=0$ uncorrelated WSM displays the expected quadratic frequency dependence in the DOS (see Fig. \ref{fig:MIT}(a)) for the 
type-I WSM considered here~\cite{Bernevig2012}. As $U$ is increased a four-peak structure characteristic of the correlated WSM emerges. Two central peaks corresponding to the renormalized WSM band dispersion together with two precusor Hubbard sidebands arise from the convolution of the spinon and rotor 
spectral functions~\cite{Florens2004,Lopez2022}. As $U$ approaches $U_c$, the simultaneous suppression of 
the quasiparticle weight and of the renormalization factor signalling the destruction of coherent quasiparticles leads to the collapse of both the height and width of the coherent contribution to the DOS. On the other hand, the weight of the incoherent Hubbard bands increase until the charge gap opens up at $U_c$. 
This is in contrast to the DMFT prediction by which only the width of the coherent contribution collapses at the Mott transition~\cite{Hofstetter2021}. 
Our results are instead consistent with cluster DMFT approaches which take into account the short-range correlations~\cite{Liebsch2011}. In the WMI phase only the Hubbard bands located around $\sim \pm U/2$ survive as in a conventional MI.

The BWSM phase arising at finite-$T$ is characterized by displaying two bands with no gap. The destruction of the coherent contribution is evident when comparing the $T=0$ DOS at $U=1$ of Fig.~\ref{fig:MIT}(a) at $T=0.02$ with the DOS of Fig.~\ref{fig:MIT}(b) in which the central coherent peaks are completely washed out. The DOS of the BWSM at finite-$T$ resembles that of the $T=0$ WMI at $U_c$, {\it i. e.}, a gapless system in which coherence is lost since $Z=0$. On the other hand the optical conductivity of the BWSM $\sigma(\nu)$ has no activation energy ($\Delta_X=0$) in contrast to the WMI but a single incoherent optical absorption band as in the WMI (see Fig.~\ref{fig:MIT}(c)). This is in contrast to the optical conductivity of conventional undoped WSMs which, at finite $T$, would display a Drude peak at zero frequency~\cite{Carbotte2016}.

Apart from transport experiments, ARPES can provide further evidence supporting the existence of the BWSM~\cite{suppl}.
ARPES in the pyrochlore iridate, Nd$_2$Ir$_2$O$_7$, located at the metallic side of the Mott insulator transition, detects quasiparticle bands consistent with the spectral function of our correlated WSM (see Fig.~\ref{fig:spectral})~\cite{Shin2016}. In contrast, the BWSM spectral function is characterized by two incoherent bands touching at the Fermi energy. 
ARPES experiments on A$_2$Ir$_2$O$_7$ with A=Eu at temperatures above the magnetic transition can corroborate the existence of our predicted intermediate BWSM.

The detection of bulk spinons present in the WMI is difficult since they are
neutral particles. However, they can be singled out through thermal conductivity $\kappa_{xx}$ experiments. Indeed, they would lead to a $T^2$ dependence: $\kappa_{xx}(T)= { 56 \pi^2 \over 5 \hbar} {k_B \over \alpha^2 \hbar {\tilde v_F}} k_B^2 T^2,$ where ${\tilde v_F}$ is the renormalized spinon velocity at the 8 Weyl nodes assuming Coulomb interactions as the scattering mechanism between spinons~\cite{suppl}.

Further evidence of spinons can come from the topological 
surface states associated with their nodal Weyl node structure 
retained in the WMI and/or BWSM. We explore such spinon surface states 
by cutting a finite slab of the crystal in the $(110)$ 
direction preserving the full periodicity in the $z$-direction 
(see Fig. \ref{fig:Fermiarcs}(a)).  
Along this cut straight Fermi arcs connecting 4 doubly degenerate 
projected Weyl points occur on the $(110)$ Surface Brillouin Zone (SBZ)
as shown in Fig. \ref{fig:Fermiarcs}(b)). 
Since the effective topological charges along this direction  are
$\nu=\pm 2$~\cite{suppl},
the Fermi arcs are doubly degenerate. 
In contrast, no Fermi arcs arise on the $(100)$ faces of 
our WMI due to a cancellation of the charges associated with the projected
Weyl nodes on the $(100)$ SBZ~\cite{Hasan2017,Goikoetxea2020}.




Despite the neutral character of the spinons, signatures of the spinon Fermi arcs arise in the surface electronic structure of the WMI. 
The electron spectral function at the $(110)$ faces $A^{R/L}({\bf k}_{||},\omega)$ shows two incoherent bands whose weights are modulated by the spinon Fermi arcs (see Fig.~\ref{fig:Fermiarcs} (c) and (d)). While $A^{R}({\bf k}_{||},\omega)$ is enhanced around the $\Gamma_s$-point, the largest intensity in $A^{L}({\bf k}_{||},\omega)$ occurs close to the SBZ edges. The characteristic intensity pattern imprinted on the Hubbard bands can be associated with the Fermi arcs at the $R/L$ surfaces shown in Fig.~\ref{fig:Fermiarcs}(b). The closing of the gap on, say, the right $(110)$ face is evident from $A^R(\Gamma_s,\omega)$ shown in Fig.~\ref{fig:Fermiarcs}(d). While the bulk spectral function shows a Mott gap, at the surface (semi)-metallic behavior arises due to the contribution of surface spinon bands at the Fermi energy~\cite{suppl}. Indeed, the largest contribution to $A^R(\Gamma_s,\omega)$ when injecting an electron at the surface comes from the simultaneous creation of a doublon at the upper Hubbard band and a surface spinon at the Fermi energy. 

The reduction of the gap at the WMI surface is in contrast to the behavior observed in conventional Mott insulators in which a reduction on the bandwidth of the 
Hubbard bands together with an {\it enhancement} of the charge gap occurs~\cite{Eckstein2013}. 
This is indeed the behavior we find in the spectral density at the $(100)$ surface which does not host spinon Fermi arcs as shown in Fig. \ref{fig:Fermiarcs}~(e).
Hence, by ARPES. We can also expect similar magnetic field effects in the WMI as found in the Mott insulator with the spinon Fermi surface\cite{Lee2023a,Lee2023b}.

Our model hosts magnetic order which is quite different from the N\'eel antiferromagnetic order expected in a Hubbard model on a simple cubic lattice. Our analysis indicates that the effective spin model is a ferromagnetic Heisenberg plus antiferromagnetic (AFM) 90$^0$ compass
model on a simple cubic lattice \cite{suppl}:
\begin{equation}
H_{\text{HFM+90$^0$compass}}= J\sum_{\langle i j \rangle, a}  ( -{\bf S}_i \cdot {\bf S}_j + 2 S_i^a S_j^a ),
\label{eq:spinmodel}
\end{equation}
with super-exchange couplings: $J= 4 t^2/U >0$. We find that the ground state of model~(\ref{eq:spinmodel}) is the layered AFM shown in Fig.~\ref{fig:phased}(a)~\cite{suppl}. The pure 90$^0$ compass model obtained by neglecting the Heisenberg term is highly frustrated since two neighbor spins along any of the $a$-axes will tend to align along that $a$-direction. Thus, there is no preferred orientation between two neighbor spins under pure 90$^0$ compass interactions. In this case, a quantum spin liquid of the type of the Kitaev spin liquid found on the honeycomb lattice can arise.


A key prediction of our work is the presence of spinon Fermi arcs at certain surfaces of the WMI. Their origin relies on the non-local Coulomb correlations in the Mott insulator which are missed by local theories such as DMFT. ARPES experiments in the WMI should detect a Mott gap in the bulk together with a gap suppression at the surfaces hosting spinon Fermi arcs. Based on our analysis, we interpret Eu$_2$Ir$_2$O$_7$ and Nd$_2$Ir$_2$O$_7$ pyrochlores above the magnetic ordering
temperature as metals with no quasiparticles, BWSMs displaying characteristic incoherent bands and no gap in ARPES. 
The WMI typically hosts unconventional magnetic order due to compass exchange couplings associated with SOC which can lead to a QSL in 3D which deserves future exploration.

\begin{acknowledgments}
	We acknowledge financial support from (Grant No. PID2022-139995NB-I00) MINECO/FEDER, Uni\'on Europea, from the Mar\'ia de Maeztu Programme for Units of Excellence in R\&D 
	(Grant No. CEX2018-000805-M). I.G.-E. acknowledges financial support from the Spanish Ministry for Science, Innovation, and Universities through FPU grant AP-2018-02748. 
\end{acknowledgments}

\bibliographystyle{apsrev4-1}
\bibliography{biblio} 
 \end{document}


\bibliographystyle{prsty}

\title{Supplementary material for Emergence of Spinon Fermi Arcs in the Weyl-Mott Metal-Insulator Transition}
\author{Manuel Fern\'andez L\'opez}
\affiliation{Departamento de F\'isica Te\'orica de la Materia Condensada, Condensed Matter Physics Center (IFIMAC) and
Instituto Nicol\'as Cabrera, Universidad Aut\'onoma de Madrid, Madrid 28049, Spain}
\author{Iñaki García-Elcano}
\affiliation{Departamento de F\'isica Te\'orica de la Materia Condensada, Condensed Matter Physics Center (IFIMAC) and
Instituto Nicol\'as Cabrera, Universidad Aut\'onoma de Madrid, Madrid 28049, Spain}
\author{Jorge Bravo-Abad}
\affiliation{Departamento de F\'isica Te\'orica de la Materia Condensada, Condensed Matter Physics Center (IFIMAC) and
Instituto Nicol\'as Cabrera, Universidad Aut\'onoma de Madrid, Madrid 28049, Spain}
\author{Jaime Merino}
\affiliation{Departamento de F\'isica Te\'orica de la Materia Condensada, Condensed Matter Physics Center (IFIMAC) and
Instituto Nicol\'as Cabrera, Universidad Aut\'onoma de Madrid, Madrid 28049, Spain}
 \date{\today}
\maketitle
 
\section{Slave-rotor approach to the Weyl-Hubbard model}
We solve the Weyl-Hubbard model using the slave-rotor mean-field 
theory (SRMFT) approach \cite{Florens2004,Florens2002} which effectively leads
to two coupled spinon and rotor hamiltonians, 
Eqs. (2) and (3) of the main text. The finite-$T$ spinon and rotor dynamics is encoded in their corresponding Matsubara Green functions reading
\begin{eqnarray}
G_{X}^{-1}({\bf k},i\nu_n) &=&\nu_n^2/U+\lambda+\epsilon_X({\bf k})
\nonumber \\
G_{f\mu}^{-1}({\bf k},i\omega_n) &=& i\omega_n-\epsilon_f^{\mu}({\bf k}) 
\label{eq::Greens}
\end{eqnarray}
where $\mu=1,2$ is the band index and $\lambda$ is the Lagrange multiplier associated with the constraint
\begin{equation}
{1 \over N}\sum_i\langle X_i^*X_i\rangle=1,
\end{equation}
which imposes charge conservation at each site $i$
with $N$ the total number of sites in the lattice. 
Coulomb interaction between electrons lead to rotor (spinon) dispersions $\epsilon_X({\bf k})$ ($\epsilon^{\alpha}_f({\bf k})$) which are renormalized by the factors $Q_X^a(Q_f^a )$ given by:
\begin{eqnarray}
Q_X^a &\equiv&\sum_{\alpha\beta}\langle f_{i\alpha}^\dagger\sigma_{\alpha\beta}^af_{j\beta}\rangle_a,
\nonumber \\
Q_f^a &\equiv&\langle X_i^*X_j\rangle_a.
\end{eqnarray}
where $a=x,y,z$ represents the three n.n. bonds 
of the simple cubic lattice. 

SRMFT leads to a set of self-consistent equations for these
renormalization factors and the constraints which, after performing the Matsubara sums, read: 
\begin{widetext}
\begin{align}
1&=Z+\frac{1}{N}\sum_{\bf k\neq 0}\frac{U}{2E_X({{\bf k}})}\left[b(E_X({{\bf k}}))-b(-E_X({{\bf k}}))\right]\nonumber\\
D_XQ_f^{a}&=Z\epsilon_X(0) +\frac{1}{N}\sum_{\bf k\neq0}e^{-i{{\bf k} \cdot {\bf r}_a }}\epsilon_X({{\bf k }})\frac{U}{2E^X_{{\bf k}}}\left(b(E_X({{\bf k}}))-b(-E_X({{\bf k}}))\right)\nonumber\\
Q_X^a&=\frac{1}{N}\sum_{\sigma}\sum_{\mu \bf k}e^{-i{\bf k }\cdot {\bf r}_a} |\chi^\sigma_\mu({\bf k}) |^2 f(\epsilon^{\mu}_f({{\bf k}})),
\label{eq::self-consistent}
\end{align}
\end{widetext}
where $D_X$ is the rotor bandwidth and $E_X({{\bf k}})\equiv\pm \sqrt{U(\lambda+\epsilon_X({{\bf k}}) )}$ are the energies of the effective quantum harmonic oscillator described by $G_X({\bf k}, i\omega_n)$ \cite{Florens2004}. The matrix elements describing the change of basis from band to spin basis reads: $\chi^{\sigma}_{\mu}({{\bf k }})= \langle {\bf k}, \sigma | {\bf k}, \mu \rangle$. 
\begin{figure}[b!]
    \centering
    \includegraphics[width=8.5cm,clip]{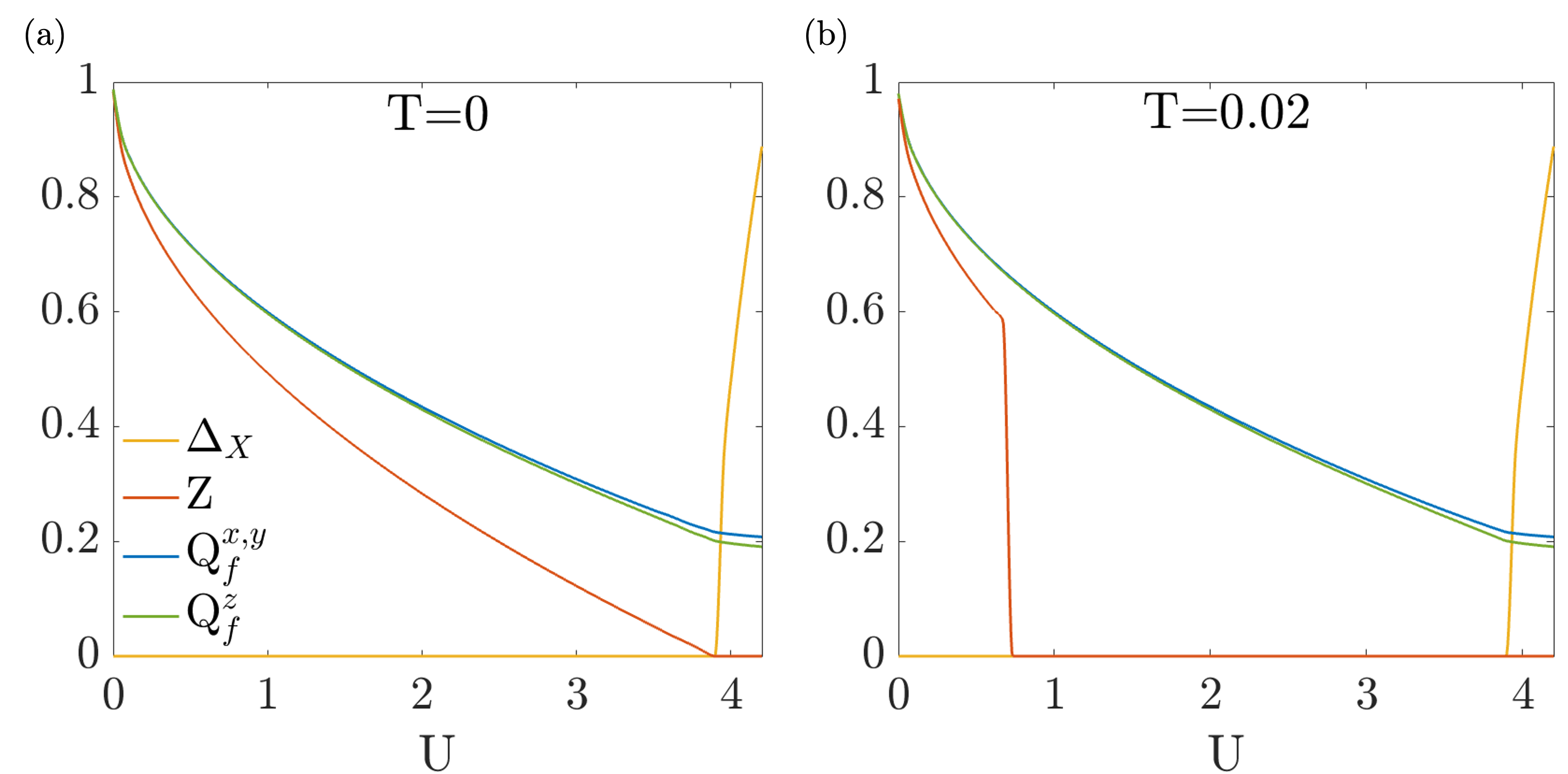} 
    \caption{Dependence of the SRMFT parameters on $U$ at
    (a) T=0 and (b) T=0.02. We have fixed $t =1$. }
\label{fig:SRparameters}
\end{figure}
Metallic phases are captured by solutions of the self-consistent equations in which Bose-Einstein condensation of the rotors occurs. This is dealt numerically by isolating the ${\bf k}=0$ mode in \eqref{eq::self-consistent}, and evaluating the 
fraction of the rotors which have condensed or the
quasiparticle weight $Z$. When 
$Z \neq 0$ the metallic state is a Fermi liquid since it contains Landau quasiparticles. The other SRMFT parameters, such as $\lambda$, $Q_X^a$ and $Q_f^a$, are iteratively determined until convergence. \cite{Lopez2022} Another key parameter for the description of the Mott insulator is the charge gap $\Delta_X$ which is given by:
\begin{equation}
\Delta_X=\sqrt{U(\lambda+\epsilon_X({{\bf k}})_{min})}.
\end{equation}

The dependence of SRMFT parameters on $U$ is shown in Fig. \ref{fig:SRparameters} both at zero and finite $T$. The single continuous Weyl-Mott metal-insulator transition at $T=0$ becomes a two-step process at $T \neq 0$: the quasiparticle weight $Z$ suppression is followed by a charge gap opening, $\Delta_X \neq 0$. We identify the intermediate gapless phase with no quasiparticles as a bad Weyl semimetal (BWSM).

\section{Electron spectral density and optical conductivity}

Single-electron properties can be characterized by computing the electron spectral density, $A^\sigma({\bf k}, \omega)$ for a given spin $\sigma$ within SRMFT: 
\begin{align}
&A^\sigma({\bf k},\omega)=\nonumber\\
&-\sum_\mu|\chi^{\sigma}_\mu({\bf k})|^2\int { d{{\bf k '}} \over (2 \pi)^3}\int d\omega' \rho^\mu_f({\bf k},\omega')\rho_X({{\bf k}-{\bf k'}},\omega-\omega')\nonumber\\
&\times\left[b(\omega-\omega')+f(-\omega')\right],
\label{eq::resolvedDOS}
\end{align}
where:
\begin{align}
    \rho_{f}^\mu({\bf k},\omega)=-\frac{1}{\pi} \text{Im} G_{f\mu}({\bf k}, \omega+i0^+)\\
    \rho_{X}({\bf k},\omega)=-\frac{1}{\pi} \text{Im} G_{X}({\bf k}, \omega+i0^+).
\end{align}

The above expressions  
\cite{Florens2004,Lopez2022} are, in principle, valid 
at any temperature. 
In phases in which there is a rotor condensed fraction, ${\it i. e.}$ $Z \neq 0$, 
$A^\sigma({\bf k},\omega)$ contains a quasiparticle contribution:
\begin{align}
    A^\sigma_{coh}({\bf k},\omega)=Z\sum_\mu|\chi^{\sigma}_\mu({\bf k})|^2\delta(\omega-\epsilon_f^{\mu}({{\bf k}})).
\end{align}
The absence of this contribution, $Z=0$, means quasiparticles do not exist and the system is incoherent. 

For simplicity, we define spin averaged quantities:
\begin{eqnarray}
A({\bf k}, \omega) &=& {1 \over 2} \sum_\sigma A^\sigma({\bf k}, \omega)
\nonumber \\
\rho(\omega) &=& {1 \over 2} \sum_\sigma \int d{\bf k} A^\sigma({\bf k},\omega).
\end{eqnarray}
These $\rho(\omega)$ and $A({\bf k}, \omega)$ are the ones shown in Fig. 2 and Fig. 3 of the main text. 

The optical conductivity shown in Fig. 2 of the main text 
is evaluated using an approximation to the Kubo formula using the electron spectral functions obtained in \eqref{eq::resolvedDOS}:
\begin{align}
\sigma({\nu}) \propto &
\int { {d{\bf k } \over (2 \pi)^3}} v_F({\bf k})^2\int d\omega A({\bf k},\omega)A({\bf k},\omega+\nu)\nonumber\\
&\times\frac{ f(\omega+\nu)-f(\omega)}{\nu},
\label{eq::sigmad}
\end{align}
where $v_F(\bf k)$ is the spinon velocity at the Weyl nodes. In this expression, the off-diagonal spin-flip terms are neglected 
since we are only interested in 
the optical gap well inside the 
Weyl-Mott phase.


\section{'Bad' Weyl semimetals in pyrochlore iridates}

The pyrochlore iridates A$_2$Ir$_2$O$_7$ display an insulating to metal transition as 
$A$ is changed from $A=Y$ (insulating) to $Pr$ (metallic). The radio of $A$ increases from 
Y to Pr. Indeed, the resistivity of Y$_2$Ir$_2$O$_7$ displays insulating behavior, while in Pr$_2$Ir$_2$O$_7$ the resistivity decreases with lowering temperature. 
A magnetic transition to an all-int all-out AFM configuration  occurs below a critical
temperature T$_c$ which varies from $T_c \sim 150 $ K for A=Y to $T_c \sim 32$ K  
for A=Nd. No magnetic transition is observed for the metallic A=Pr down to the 
lowest temperatures.

As noted previously, Y$_2$Ir$_2$O$_7$ displays absolute values of the resistivity consistent with 'bad' Weyl semi-metallic behavior. This means that the Mott-Ioffe-Regel limit is violated, $\rho^{expt} >>  \rho^D= {3 \pi^2 \hbar \over e^2 N (k_T^2 l) } $ where $N$ is the number of Weyl nodes and
$k_T = {k_B T \over \hbar v_F}$, is a thermal Fermi vector characterising thermal
excitations around the Weyl nodes (of the undoped system). 
For the observed resistivities in the $Y$ compound $\rho^{expt} \sim  10^7-10^2$ $\Omega$cm
at $T=3-60$ K implying mean-free paths $k_T l << 2 \pi$ 
clearly violating the MIR limit. This implies 'bad' Weyl semimetal behavior. 
The MIR limit is violated even at higher
temperatures at which metallic behavior is observed. 
For Y($\rho^{expt}(T= 200K) \sim 10 \Omega$cm 
$Nd$($\rho^{expt}(T= 200K) \sim 5 \times 10^{-3}$ $\Omega$cm ) and Eu ($\rho^{expt}(T= 200K) \sim  30 \times 10^{-3}$ $\Omega$cm ) are consistent with $ k_T l < 2 \pi $. 
Finally the Pr iridate($\rho^{expt}(T= 200K) \sim 2 \times 10^{-3}$ $\Omega$cm) has the lowest 
resistivity consistent with metallic behavior.
Hence,  the 'bad' metallic behavior observed correlates well with the 
strength of the Coulomb interactions. As the Coulomb interaction is increased 
by decreasing the radii of the A ions, the resistivities observed become large
leading to mean-free paths violating the MIR limit. This occurs even at temperatures
above the magnetic transition. We conclude that electronic correlations are 
behind the bad metallic behavior observed. 

\section{Thermal conductivity}
    
The Weyl-Mott insulator has a charge gap so no DC currents can circulate in the system. In contrast, 
spinons are gapless but since they are neutral particles they can carry heat but no charge. 
Hence, we can expect a non-zero contribution to the longitudinal thermal conductivity, $\kappa_{xx}$, 
from the spinons in the WMI phase. The electron-electron interaction, $U$, induces a finite lifetime, $\tau$, on the spinons not included in SRMFT whose dispersion is that of the WSM 
renormalized by the Coulomb interaction. 

The transition rate solely due to Coulomb repulsion reads\cite{Burkov2011}:
\begin{equation}
    {1 \over \tau} \sim  \alpha^2 
    {\text max } \{ {\omega \over \hbar}, { k_B T \over \hbar} \},
    \label{eq:tau}
\end{equation}
where $\alpha= {e^2 \over  \epsilon \hbar {\tilde v_F} }$  with $\epsilon({\bf k})=\hbar {\tilde v_F} |{\bf k}|$, with
$ {\tilde v_F} \approx Q_f v_F$, the renormalized spinon velocity.
The semiclassical expression (Fiete et. al.) for the thermal conductivity
in a WSM under Coulomb scattering can be expressed as:
\begin{equation}
\kappa_{xx}(T) = {7 \pi^2 \over 5 e^2} k_B^2 \sigma_{xx} T. 
\end{equation}
Using (\ref{eq:tau}) in  $\sigma_{xx}$ of a WSM with $N$ nodes under Coulomb scattering gives:  $\sigma_{xx} = {N e^2 \over \hbar} {k_B T \over \alpha^2 \hbar {\tilde v_F} } $, one finds: 
\begin{equation}
\kappa_{xx}(T)= { N 7 \pi^2 \over 5 \hbar} {k_B \over \alpha^2 \hbar {\tilde v_F}} (k_B T)^2.     
\end{equation}
Hence, we should expect a contribution,
$\kappa_{xx} (T) \propto T^2$, in the WMI solely due to spinons at low temperatures.
\section{Spinon Fermi arcs}
The WMI is a fractionalized state with gapped charge degrees of freedom but gapless magnetic excitations carried by the spinons conserving the original nodal structure of the WSM. Thus, in the WMI we expect topological surface states connecting the Weyl nodes which, unlike in conventional WSM, they carry no charge but only neutral spin excitations. We explore such unconventional spinon surface states by cutting a finite slab of the crystal in the $r_{\perp}=\hat{x}+\hat{y}$ direction retaining periodicity along 
the $r_{\parallel}=\hat{x}-\hat{y}$ and $r_z=\hat{z}$ directions. 
In this rotated basis $(r_{\perp},r_{\parallel})$, the original 8 Weyl become degenerate leading to only
4 degenerate Weyl points with topological charges of $\nu=\pm 2$. 
\begin{figure}[t!]
    \centering
    \includegraphics[width=8.5cm,clip]{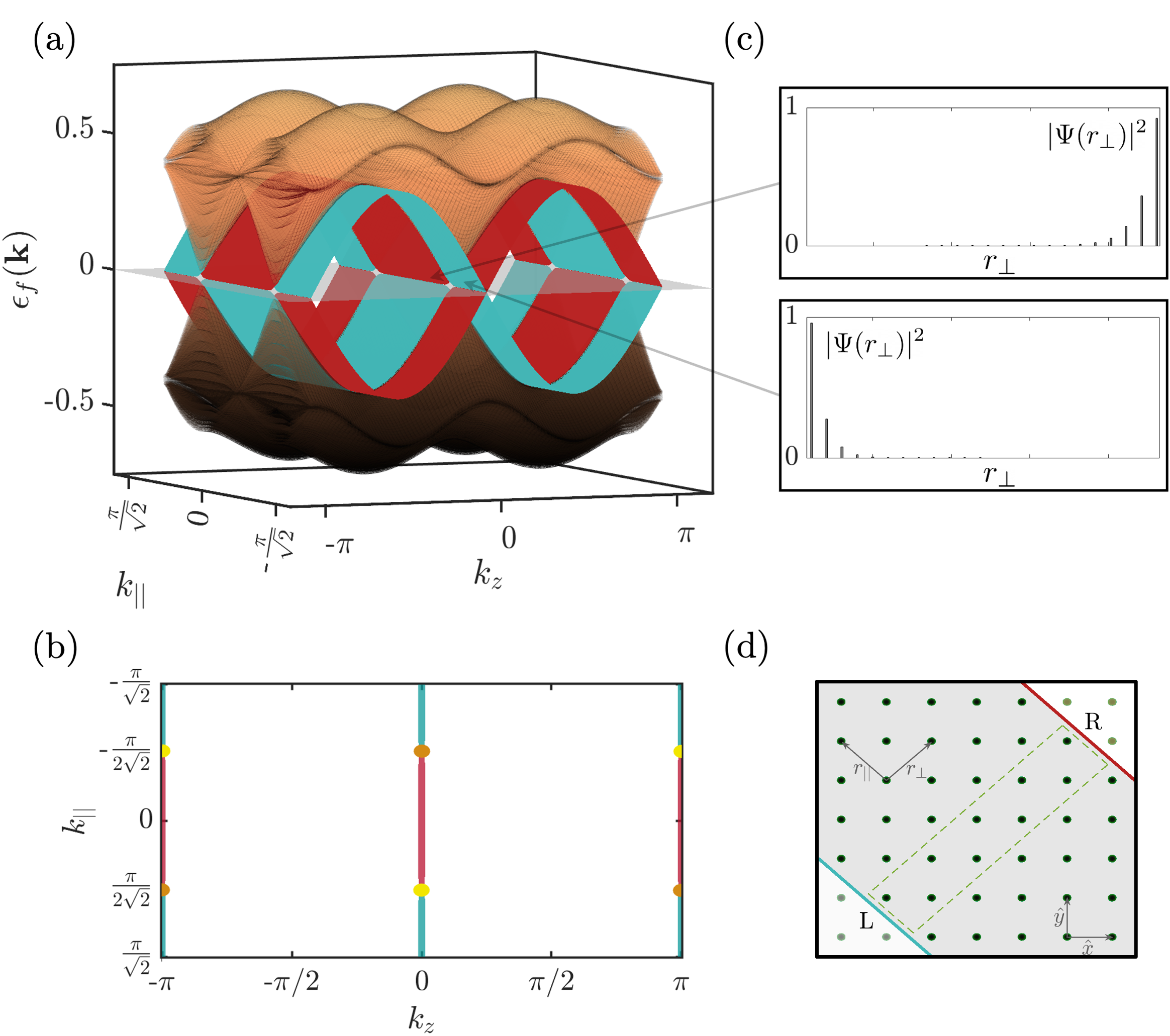}
    \caption{Neutral spinon surface states in the WMI (a) Spinon bands for a finite slab along planes in the $r_{\perp}$ direction within the Weyl-Mott insulator for $U=4.2$. 
    (b) Surface Brillouin zone (SBZ) containing both the Weyl points projections and the spin-up spinon Fermi 'arcs' 
    at left (red) and right (cyan) surfaces of the slab. 
    (c) Zero-energy eigenvalue amplitudes $|\Psi|^2$ for spin-up spinons as a function of the unit cell sites along $r_{\perp}$ for the chosen ${\bf k}$-point. 
    (d) Projection of the slab on the $x-y$ plane.
    }
    \label{fig:Fermiarcs}
\end{figure}
The spinon band structure in the WMI shown in Fig. \ref{fig:Fermiarcs}(a) contains flat spinon Fermi arcs connecting the 4 Weyl points which leads to straight Fermi arcs when projected onto the 
surface Brillouin zone (see Fig. \ref{fig:Fermiarcs}(b)).
Despite apparently closed, these Fermi arcs are, in fact, independent and open, as they should. Indeed, the spatial location of 
surface spin-up spinons (whether it propagates along the right or left sides of the slab) is different for wavevectors on consecutive Fermi arcs as shown in Fig. \ref{fig:Fermiarcs}(c)). This can be further analyzed by adding an extra mass term which splits the apparently closed Fermi arcs into separate sections as discussed in the 
next section.

\section{Mass term effect on Fermi arcs }
In this section we analyze peculiarities of the topological surface states of our particular Weyl system. Since spinons immersed in the fractionalized WMI retain the topological properties of the uncorrelated WSM electrons, we focus on the bare Weyl system.
The Fermi arcs shown in Fig. 4 of the main text instead of being apparently closed, are opened independent Fermi arcs. 

\begin{figure}[t!]
    \centering
    \includegraphics[width=8cm,clip]{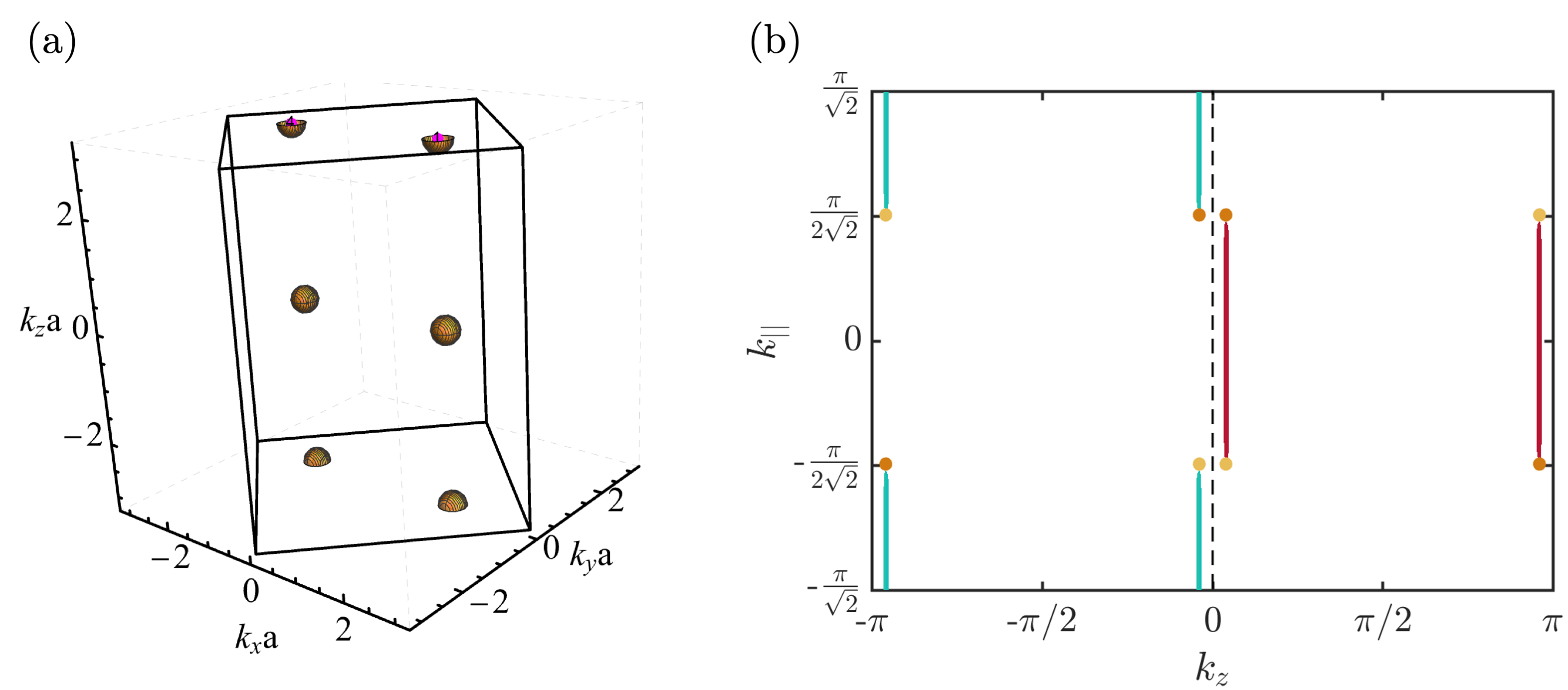} 
    \caption{(a) FBZ in the rotated basis (b) Fermi arcs with $m=0.05$.}
\label{fig:massFermi}
\end{figure}

This is what we try to elucidate here by studying the effects of a mass term added to the slab lattice. This is equivalent to add an staggered potential distinguishing odd and even sites of the slab unit cell.
This symmetry breaking originate a splitting of the 4 Weyl points projection recovering the original 8 Weyl points with single instead of double topological charges as Fig. \ref{fig:massFermi}(a) shows. The resulting straight Fermi arcs have different $k_z$ momentum and are located on opposite slab surfaces (see Fig. \ref{fig:massFermi}(b)).

\section{Slab spectral functions}

We now discuss our calculation of the spectral density in the slab of Fig. \ref{fig:Fermiarcs} (d) and that are shown in Fig. 4 of the main text. The layers along the slab in Fig. \ref{fig:Fermiarcs} (d) are labeled by $l$ with $l=1,...,M$  where $l=1$ corresponds to left (L)
amd $l=M$ to right (R) surfaces. We describe the electronic states in the slab through the set of quantum numbers ${ ({\bf k}_s, l,\sigma)} $ where ${\bf k}_s=(k_{||},k_z)$ is the momentum in a layer. The spinon Green's function in matrix form reads:
\begin{equation}
    G_f({\bf k}_s,\omega_n)=(\omega\mathbbm{1}-H_f({\bf k}_s))^{-1},
\end{equation}
where matrices are of order $2M$ and ${\bf k}_s=(k_{||},k_z)$. $H_f ({\bf k}_s)$ is the spinon hamiltonian including the renormalization factors, $Q^a_f$. 

The rotor Green's function is obtained in a slightly different way by expressing the corresponding quantum $XY$ hamiltonian as:
\begin{eqnarray}
H_X &=& \sum_{{\bf k}_s} \sum_{l} \left({\frac{U}{2}} L^2+\lambda X_l^*({\bf k}_s) X_l({\bf k}_s )\right) 
\nonumber \\
&+& tQ_X\sum_{{\bf k}_s} \sum_{l,l'} X_{l}^*({\bf k}_s)X_{l'} ({\bf k}_s)
\end{eqnarray}
where
$Q_X (Q_X\equiv Q_X^a$ for $a=x,y,z$) is the bulk rotor renormalization factors and $\lambda$ is the bulk Lagrange multiplier. While $H^{SRMF}_U$ describes the on-site Coulomb interaction on each layer within SRMFT, $H^{SRMF}_t$ is the SRMF renormalized kinetic energy connecting all sites. Using Dyson's equation we obtain the rotor Green's function of the slab which, in matrix form, reads:
\begin{equation}
    G_{X}({\bf k}_s,\nu_n)=\left(\mathbbm{1}-G_{X}^{(0)}({\bf k}_s, \nu_n)\Sigma({\bf k}_s)\right)^{-1}G_{X}^{(0)}(\nu),
\end{equation}
where matrices are of order $M$ since rotors don't carry spin. The rotor Green's function reads:
\begin{equation}
G_{X,ll'}^{(0)}({\bf k}_s,\nu_n)^{-1}=({\nu_{n}^2 \over U}+\lambda) \delta_{ll'},
\end{equation}
which is independent of ${\bf k}_s$ since it is local, 
and the self-energy:
\begin{equation}
\Sigma_{ll'}({\bf k}_s)=t Q_X X^*_l({\bf k}_s)X_{l'}({\bf k}_s). 
\end{equation}

The spinon and rotor spectral functions can be obtained from the diagonal elements of the slab Greens functions as
\begin{align}
\rho_{f}^{l\sigma}({\bf k}_s,\omega)=-{ 1 \over \pi} \mathfrak{Im}G_{f}^{l\sigma}({\bf k}_s,\omega+i\eta),\\
\rho_{X}^{l}({\bf k}_s,\omega)=-{ 1 \over \pi} \mathfrak{Im}G_{X}^{l}({\bf k}_s,\omega+i\eta).
\end{align}
where we have defined: $G_{f}^{l\sigma,l\sigma}\equiv G_{f}^{l\sigma}$ and $G_{X}^{l,l}\equiv G_{X}^{l}$.

The physical electron spectral function at layer, $l$, of the slab $A^{l\sigma}({\bf k}_s,\omega)$ is obtained from:
\begin{align}
    A^{l\sigma}({\bf k}_s,\omega)=&-\int { d{\bf k'} \over (2 \pi)^2} \int d\omega'\rho_f^{l\sigma}({\bf k}_s,\omega')\rho_X^l({\bf k}_s-{\bf k'}_s,\omega-\omega')\nonumber\\
    &\times [b(\omega-\omega')+f(-\omega')].
    \label{eq::Adslab}
\end{align}
From the above expression we obtain $A^{l\sigma}({\bf k}_s,\omega)$ 
at the $L(R)$ surface of a slab, $l=1(M)$, or in 
the bulk, $l=M/2$, shown in Fig. 4 of the main text. 
For simplicity, we focus on the spectral function of spin-up electrons, $\sigma=\uparrow$, so we define $\rho^{l}_{f/X}\equiv \rho^{l\uparrow}_{f/X}({\bf k}_s,\omega)$ and $A^{l}({\bf k}_s, \omega)\equiv A^{l\uparrow}({\bf k}_s,\omega)$. 

\begin{figure}[t!]
    \centering
    \includegraphics[width=9cm,clip]{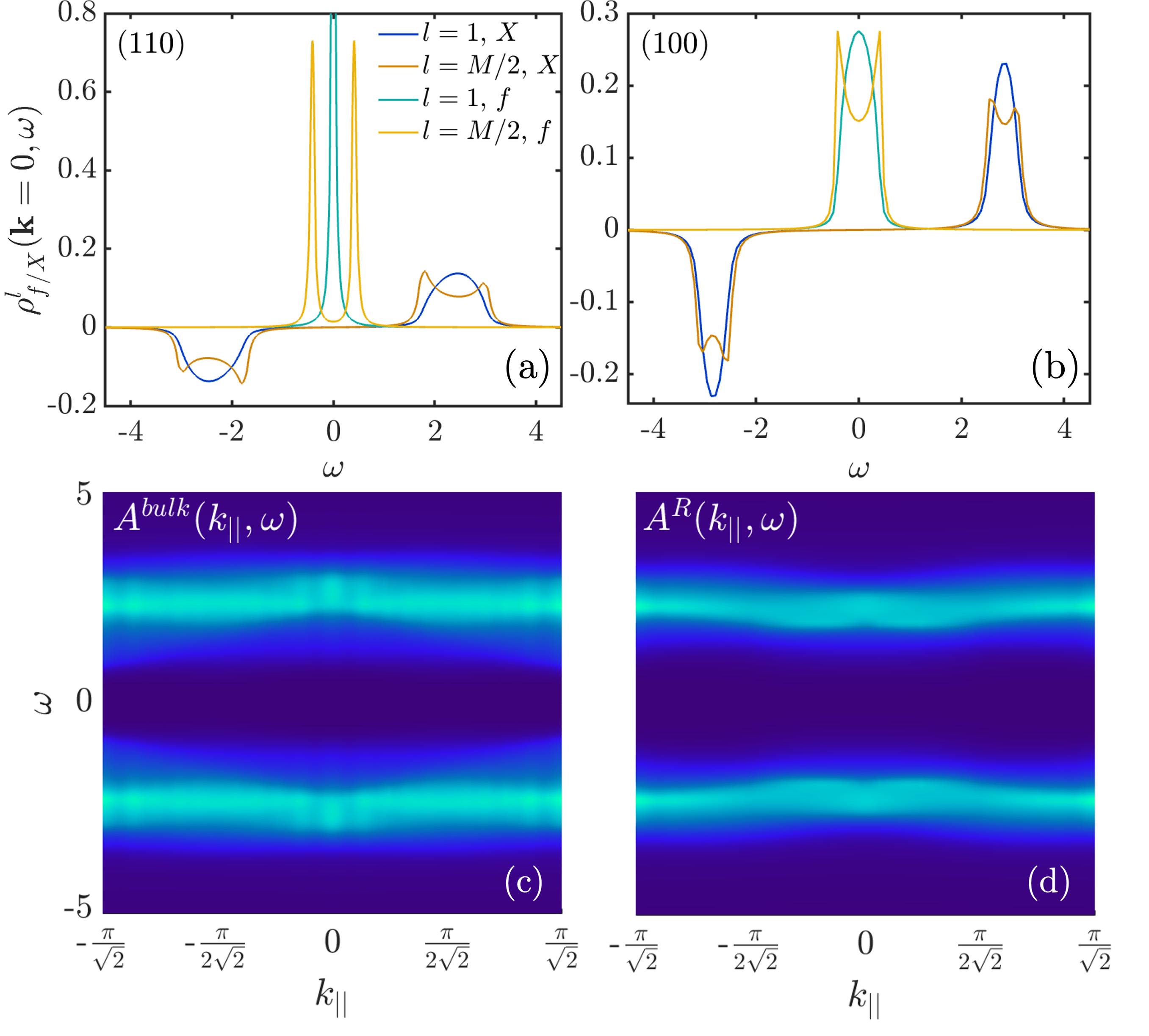} 
    \caption{
    Spinon (rotor) spectral densities, $\rho^l_f$ ($\rho^l_X$) at ${\bf k}_s=0$ evaluated at the bulk ($l=M/2$) and surface ($l=1$) of a (a) $(110)$ and (b) 
    $(100)$ slab. The slab spectral function along the ${\bf k}_s=(k_z=0,k_{||})$ direction 
    in the (c) bulk, $A^{bulk}({\bf k}_s,\omega)$ and (d) at the R-surface, $A^{bulk}({\bf k}_s,\omega)$, of a $(110)$ slab. We have taken $U=3.9$ and 
    $T=0.01$. We fix $t \equiv 1$. }
\label{fig:Ad}
\end{figure}

The spinon and rotor spectral functions of two different slabs are shown in Fig. \ref{fig:Ad}(a),(b) in the WMI. 
While the $(100)$ slab is topologically trivial, the $(110)$ hosts surface spinon Fermi arcs as discussed above. This leads to quite different spinon spectral functions at the surfaces of the $(110)$ vs. $(100)$ slabs. While the $(100)$ surface displays semi-metallic behaviour, a large zero-energy weight 
occurs at the $(110)$ surface which is associated with the topological 
surface bands traversing the Mott gap of the WMI. Such large spinon spectral density leads to the gap suppression in the electron spectral density 
$A({\bf k}_s=0,\omega)$ shown in Fig. 4 (d) of the main text. In contrast, an enhancement of the gap occurs at the $(100)$ surface. The effect of spinon Fermi arcs on the electron spectral function is evident when comparing electron spectral densities $A^{l\uparrow}({\bf k},\omega)$ in the bulk ($l=M/2$) and the, say, $R$-surface ($l=1$) as shown in Fig. \ref{fig:Ad}. 

\section{Magnetic model}
We perform a mapping of the original Hubbard model onto a superexchange Hamiltonian for localized spins 
in the $U\gg t$ limit. Starting from the hopping hamiltonian
between neighboring sites along the $\alpha$ direction:
\begin{equation}
H_{ij}^a= t_a e^{i \phi_a} c^\dagger_i \sigma^a c_j,
\end{equation}
with $a=x,y,z$. We can obtain the effective superexchange hamiltonian between two localized spins to $O(t^2/U)$ as:
\begin{eqnarray}
H^{(2)a} &=& -{2 t_a^2 \over U} H_{ij}^a H_{ji}^a 
\nonumber \\
&=& {4 t_a^2 \over U} \left( -{\bf S}_i  {\bf S}_j + 2 {\bf S}_i^a {\bf S}_j^a 
-n_i + { n_i n_j \over 2}  \right).
\nonumber \\
\end{eqnarray}

Neglecting irrelevant terms, the effective spin model is a Heisenberg + 90$^0$ compass model on the simple cubic lattice:
\begin{equation}
H_{\text{HFM+90$^0$compass}}=\sum_{\langle i j \rangle, \alpha} J_a ( -{\bf S}_i \cdot {\bf S}_j + 2 S_i^a S_j^a ),
\label{eq:spinmodel}
\end{equation}
The superexchange coupling constants: $J_a= 4 t_a^2/U >0$. 
Assuming isotropic exchange couplings: $J_a=J$, we obtain the spin model quoted in Eq. (4) of the main text. 

\begin{figure}[t]
    \centering
    \includegraphics[width=8.cm,clip]{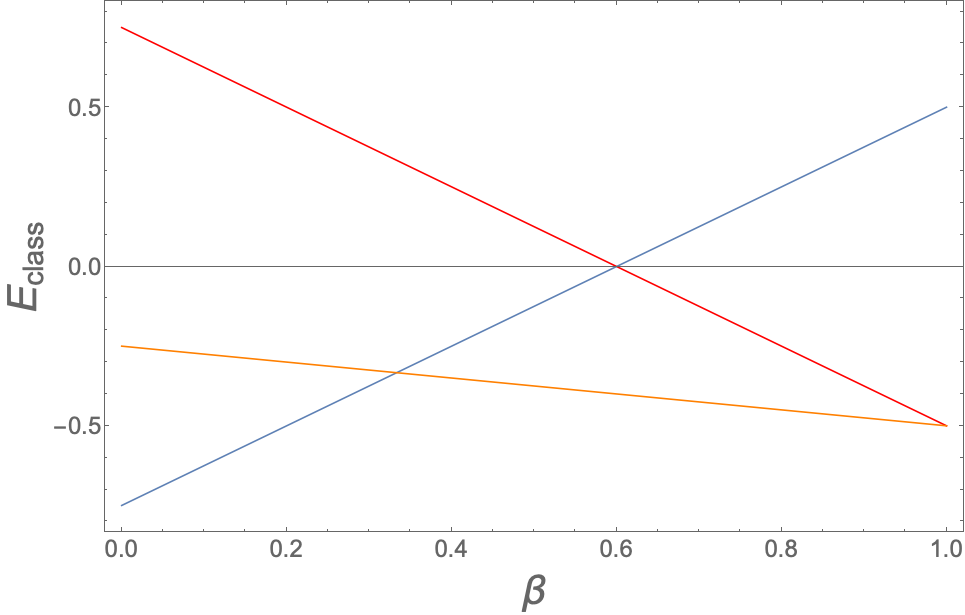}
    \caption{Classical energies of the Heisenberg-Kitaev model (\ref{eq:spinmodel}). The dependence of the FM (blue), N\'eel (red) and layered AFM (orange) energies with $\beta$ are shown. A transition from the FM to the layered AFM occurs at $\beta=1/3$.} 
    \label{fig:Eclass}
\end{figure}

We can gain insight into the magnetic order of the model by analyzing the ground state of  $H_{ij}^a$ on two sites. While the ground state of $H_{ij}^x$ is: 
\begin{equation}
| \Psi^x_0 \rangle = (-| \uparrow, \uparrow \rangle +|\downarrow, \downarrow \rangle )/\sqrt{2},
\end{equation}
with energy $E^x_0=-2 t_x^2/U$. The ground state
in the $y$-direction is:
\begin{equation}
| \Psi^y_0 \rangle = (| \uparrow, \uparrow \rangle +|\downarrow, \downarrow \rangle )/\sqrt{2},
\end{equation}
with energy $E^y_0=-2 t_ y^2/U$.
Finally the ground state of $H_{ij}^z$ with energy $E^z_0=-2 t_z^2/U$ is the triplet state: 
\begin{equation}
| \Psi^z_0 \rangle = (| \uparrow, \downarrow \rangle + | \downarrow, \uparrow \rangle)/\sqrt{2}
\end{equation}
Our analysis indicates how two neighbor spin interactions are ferromagnetic.
Possible magnetic states associated with model (\ref{eq:spinmodel}) may be inferred by exploring its extended version:
\begin{equation*}
\tilde{H}_{\text{HFM+90$^0$compass}}=-J_2\sum_{\langle i j \rangle, \alpha} {\bf S}_i \cdot {\bf S}_j 
+ J_1 \sum_{ij,a} S_i^a S_j^a.
\label{eq:HK}
\end{equation*}
where the parameterized couplings: $J_2=1-\beta$, $J_1=2 \beta$. Not that for $\beta=1/2$, $J_1=2 J_2$ we recover the original spin model (\ref{eq:spinmodel}). The dependence of the classical energies with $\beta$ is shown in Fig. \ref{fig:Eclass}. A layered-AFM phase arises at intermediate $\beta$.
The energies of the three states considered are: 
\begin{eqnarray}
E_{\text{N\'eel}}(\beta)=(3-5 \beta)/4
\nonumber \\
E_{\text{FM}}(\beta)=-(3-5 \beta)/4
\nonumber \\
E_{\text{layered-AFM}}(\beta)= -(1 + \beta)/4.
\end{eqnarray}
The layered-AFM can be understood by re-expressing $\tilde{H}_{HFM+90^0 compass}$ in a rotated spin basis, ${\bf S}_i$. 
Starting on a given site with fixed spin, 
$S_i={\tilde S}_i$, neighboring 
spins in the $x$, $y$ and $z$ directions are rotated by 
$\pi$ around the $x$, $y$ and $z$-axis, respectively. 
Our model (\ref{eq:spinmodel}) which corresponds to $\beta=1/2$, is exactly mapped onto the AFM Heisenberg model:
\begin{equation}
H={1 \over 2} \sum_{\langle i j \rangle } {\tilde {\bf S}}_i \cdot {\tilde {\bf S}}_j,
\end{equation}
whose ground state is the N\'eel state but 
with unsaturated localized spins due to quantum fluctuations
not considered in the calculation. Re-expressing the N\'eel state in the original spin basis leads to the layered AFM phase which is analogous to the zig-zag phase of the Heisenberg-Kitaev model on a honeycomb lattice. As shown in Fig. \ref{fig:Eclass} the FM to layered AFM transition occurs at $\beta=1/3$ and is stable up to $\beta=1$ where it becomes degenerate with the N\'eel state.

\bibliographystyle{apsrev4-1}
\bibliography{supp}